\documentstyle[aps] {revtex}

\begin{document}
\draft

\title{ Quantum Geometrodynamics I:\\
        Quantum--Driven Many--Fingered Time}

\author{Arkady Kheyfets\footnote{E-mail: kheyfets@odin.math.ncsu.edu}}
\address{Department of Mathematics, North
   Carolina State University, Raleigh, NC 27695-8205}
\author{Warner A.~Miller\footnote{E-mail: wam@regge.lanl.gov}}
\address{Theoretical Division (T-6, MS B288),
   Los Alamos National Laboratory,
   Los Alamos, NM 87545}

\date{May 13, 1994}

\maketitle

\begin{abstract}
The classical theory of gravity predicts its own demise --
singularities.  We therefore attempt to quantize gravitation, and
present here a new approach to the quantization of gravity wherein the
concept of time is derived by imposing the constraints as
expectation-value equations over the true dynamical degrees of freedom
of the gravitational field -- a representation of the underlying
anisotropy of space.  This self-consistent approach leads to
qualitatively different predictions than the Dirac and the ADM
quantizations, and in addition, our theory avoids the interpretational
conundrums associated with the problem of time in quantum gravity.  We
briefly describe the structure of our functional equations, and apply
our quantization technique to two examples so as to illustrate the
basic ideas of our approach.
\end{abstract}


\section{Classical Dynamics of General Relativity and its Quantization.}
\label{I}
        It is reasonable to say that very few problems of modern
theoretical physics have attracted as much attention and as much
effort as gravity quantization.  Simply stated, the classical theory
of gravity predicts its own demise.  Under quite general circumstances
it has been shown that the classical theory evolves toward
singularities --- singularities in the fabric of
spacetime.\cite{Mis69,BKL71,BK69} Secondly, the vacuum fluctuations of
spacetime, commonly referred to as spacetime foam, may exhibit
``collective mode oscillations'' (CMO's) of a rich enough variety and
spectrum so as to provide physics, once and for all, with a quantum
geometric description of all matter and field interactions.\cite{Whe57}
We are so fantastically far from solving such lofty problems;
nevertheless, these goals provide us with the motivational foundations
for our research in quantum geometrodynamics.  However, modern
attempts to quantize gravity are interwoven with contradictions,
internal inconsistencies and conceptual ambiguities.  In particular,
the following three problems have received much attention and have
plagued the development of a theory of quantum gravity: (1) the
square--root Hamiltonian problem; (2) the problem of time; and (3) the
Hilbert space problem. These difficulties are so pervasive that they
persist, and their nature remains about the same even when restricted
to the very simplest of models --- quantum cosmology with a finite amount
of degrees of freedom.

Upon a careful analysis of existing approaches to the gravity
quantization we have come to the conclusion that the source of this
state of affairs can be traced to a misinterpretation of the classical
dynamical theory of general relativity.  To appreciate this statement
one should note that every single attempt of gravity quantization is
based on the original ADM\cite{ADM62} picture of the gravitational
field dynamics. In this picture, the dynamical evolution of the
gravity field manifests itself as a change from one spacelike
3--geometry to another. In other words, the configuration space of
geometrodynamics is believed to be Wheeler's superspace.\cite{Whe70}
Such an approach does not utilize York's analysis of the gravitational
field's degrees of freedom and the initial-value
formulation.\cite{Yor72,Yor73} This is not surprising as the
foundations of quantum gravity \cite{Per62,DeW67} were originally
formulated before York completed his investigation.

The ADM analysis of Einstein equations leads to a natural split of
these equations into (1) six evolution equations, and (2) four
constraints that enforce the general covariance.\cite{MTW70} This
situation is similar, in a sense, to that of the dynamics of gauge
fields. In the case of a homogeneous cosmology, the situation closely
resembles the dynamics of a relativistic particle.\cite{RS75} The
peculiarity of the gravitational field, as described by general
relativity, is that its dynamics is maximally constrained, i.~e.,
given the constraints on all possible spacelike slices of a given
spacetime, it is possible to recover the full system of Einstein
equations.\cite{Ger69} This particular property has led some
researchers to the conclusion that the gravity field dynamics is
determined completely by the constraints.\cite{Kuc92} The last
statement should be, in our opinion, handled with extreme caution.
Its proper interpretation demands additional assumptions that are
frequently made implicitly and, if not analyzed carefully, might lead
to numerous paradoxes in both classical and quantum dynamics. A
detailed analysis of this situation has been provided by
U.~Gerlach.\cite{Ger69} When applied to gravity quantization, this
line of reasoning leads to the conclusion that, to quantize such a
fully constrained field, one needs only to quantize the constraints.
Such an approach leads to a wave equation, called the Wheeler--DeWitt
equation, which is more akin to the Klein--Gordon equation rather than
the Schr\"odinger equation.  An equation of this type, as it is well
known, cannot be interpreted in a one--particle (in case of gravity,
one--Universe) representation.

The original ADM quantization proposal was different. According to it,
one had to solve constraints before quantization and to extract a
Schr\"odinger equation from the Hamilton--Jacobi classical equation.
The peculiar feature of this approach was the emergence of the so
called square--root Hamiltonian problem.\cite{RS75,Mis69} This
difficulty has, to our knowledge, never been resolved.

Both the ADM and Dirac approaches in any formulation (canonical, path
integrals, Euclidean path integrals, etc.) lead to difficulties of
about the same nature, which were captured so dramatically by
K.~Kucha\v r when he reformulated them as ``the problem of
time.''\cite{Kuc92} Roughly speaking, both approaches eliminate a
possibility of including in the theory a natural concept of time or an
observer (cf.~also a discussion of the issue by W.~Unruh\cite{Unr94}).
We prefer a slightly different formulation. In our opinion, the source
of such difficulties arises when one treats the whole 3--geometry of
spacelike slices as dynamical and quantizes the entire 3--geometry.
Mathematically, it is expressed via imposing the commutation
relations\cite{DeW67} on all the components of the 3--metric. One
should keep in mind, however, that in general relativity the system
described by the 3--metric or even the 3--geometry includes in itself
the observer and his clock. In standard quantum mechanics, or even in
the quantum field theory, an observer is external with respect to the
quantum system.  The observer is classical and has an external
classical clock.  There cannot be an external observer in the
description of the gravitational field because the gravitational
system is the Universe itself. Both ADM and Dirac's approaches
essentially quantize the observer and his clock on equal footing with
the rest of the system (J.~A.~Wheeler would say that many--fingered
time is directly quantized).  Whether such a quantized observer and
his clock can function in a fashion providing an opportunity to
describe the system consistently is not clear. A discussion of such a
possibility, however exciting, clearly would lead us far beyond the
scope of this paper. We only wish to mention here that this difficulty
has been noticed by some researchers, most notably by Gell--Mann and
Hartle and has led them to propose a generalized form of quantum
theory based on the ideas of histories and decoherence
functionals.\cite{Har92}. What we propose here is quite a different in
nature than the Gell-Mann--Hartle theory.

\section{Quantum Geometrodynamics: Basic Ideas.}
\label{II}

We propose an alternative approach to the quantization of gravity.
Our approach is based on the post-ADM achievements made in classical
geometrodynamics.\cite{Whe88} In particular, we are referring to
York's solution of the initial--value problem and his analysis of the
gravitational degrees of freedom.\cite{Yor72,Yor73} This development
was initially motivated by Wheeler's semi--intuitive remark that the
3--geometry of a spacelike hypersurface has encoded within it the two
gravitational degrees of freedom as well as its temporal location
within spacetime.  It is this notion that the 3--geometry is a carrier
of information on time that has been referred to as ``Wheeler's {\sl
many--fingered time}.''\cite{BSW62,BSW63,MTW70} It was J. York who
first made this thesis precise.  He forwarded what has now become
almost the canonical split of the 3--geometry into its underlying
conformal equivalence class (its {\sl shape} representing the two
dynamical degrees of freedom of the gravitational field coordinate per
space point) and the conformal scale factor (its {\sl scale}
representing Wheeler's many--fingered time).  Only the conformal
3--geometry is dynamical in the sense that it can be specified freely
as the initial data. The scale factor is nondynamical and essentially
specifies Wheeler's many--fingered time. It is determined by both a
slicing condition (the fixation of a field of observers) and the
constraints (enforcing general covariance).  The results of York have
demonstrated that the true dynamical part of the gravitational field
is not the 3--geometry but only its conformal part, and that the
proper configuration space or ``arena for geometrodynamics'' should be
the underlying conformal superspace rather than Wheeler's superspace.
The conformal scale factor, York's representation of Wheeler's
many--fingered time, thus becomes an external parameter and should be
explicitly treated as such in any viable quantization scheme.

None of the existing attempts of gravity quantization truly adopts
York's analysis of the gravitational degrees of freedom. We propose,
on the contrary, to design a consistent quantization procedure by
taking York's construction as a~priori. In the classical theory, we
start from the standard Lagrangian and the associated action (with
appropriate boundary terms as needed), develop the standard
variational dynamical picture over conformal superspace and treat the
scale factor (and, in more general setting, the coordinatization
parameters) as an external parameter. It is clear that the Hamiltonian
of such a theory will not be the usual super-Hamiltonian.
Nevertheless, one can develop an entire dynamical picture based on
this Hamiltonian by deriving either the Hamiltonian equations or the
Hamilton--Jacobi equation. However, these equations are incomplete.
They contain as yet unknown functions related to the scale parameter.
One can complete the system of equations by adding to the Hamilton
equations, or to the Hamilton--Jacobi equation, the standard
constraint equations of general relativity (they cannot be derived
from variational principles in such a theory). The resulting equations
are equivalent to the standard equations of classical
geometrodynamics.

For the purpose of quantization, we start from our Hamilton--Jacobi
equation, describing effectively what we refer to as {\sl conformal
geometrodynamics} (evolution of the dynamical variables corresponding
to the conformal part of the 3--geometry) in a scale parameter
dependent external field. We augment this with the four constraint
equations so as to recover the relationship between the scale
parameter and the dynamical variables. Using the Hamilton--Jacobi
equation we write down the Schr\"odinger equation treating the scale
parameter as an external classical field and quantizing only the true
dynamical variables. This Schr\"odinger equation can be solved (cf.,
for instance the example of the Bianchi~1A cosmological model in
Sec.\ref{IV}). The solution will ordinarily depend on the
many--fingered time parameter (as yet unspecified) as well as on the
coordinatization parameters.  To specify these functions we use the
constraint equations. The procedure for this follows from our
interpretation of the constraint equations.  Whereas the approach of
ADM attempted to isolate the dynamical degrees of freedom of the
gravitational field by imposing the constraints at a classical level
prior to quantization, we propose here a weaker condition that the
four constraints be imposed only on the expectation values of the
conformal dynamics.  That is, we impose the constraint equations as
expectation--value equations using the wave functional obtained from
our Hamiltonian.  In so doing we explicitly avoid the interpretational
conundrums associated with the problem of time as well as square--root
Hamiltonians, and we form a ``classical'' gravitational clock driven
by the quantized geometrodynamic system --- i.e. {\sl quantum--driven
many--fingered time}.

The goal of this paper is to clarify our basic thesis regarding the
quantization of gravity --- quantum geometrodynamics. As we have
emphasized, our idea implies a change in the quantization procedure of
constrained dynamical systems. Such a procedure can be applied to
other more familiar constrained Hamiltonian systems.  In order to
illustrate the salient features of our novel approach we cannot think
of a more suitable example than the relativistic particle.  Therefore,
in the next section we apply our quantization procedure to a free
relativistic particle. We do not discuss the issue of its usefulness
for particle dynamics, as this issue should be a topic of a separate
research project. We use this example merely as a simple illustration
of our approach and a test of our procedure. The utility of this
example lies entirely with its simplicity, clarity in illustrating
our procedure, and in demonstrating that this procedure can be used in
geometrodynamics.

\section{Time as an external field in the quantum dynamical picture
         of a free relativistic particle.}
\label{III}

We start from the standard dynamics of a particle in the
super-Hamiltonian\cite{MTW70} formulation and recover its Lagrangian.
Then we change the dynamical picture via removing the time coordinate
out of the set of the dynamical variables. This results in a new
Hamiltonian. The time coordinate and its functions are treated as
external fields. This procedure leads to an incomplete system of the
equations of motion. The system is completed via imposing additionally
the super-Hamiltonian constraint --- a constraint that does not
follow from the equations of motion in this new approach. It is
considered merely as a relation between the dynamical variables and
the external field. The resulting description is equivalent to the
standard one within the classical theory, but leads to a considerably
different quantum theory.

The super-Hamiltonian for a free particle is given by
\begin{equation}
\label{1}
{\cal H} = {1\over 2m}\left( m^2 + p^2\right) = {1\over 2m}\left( m^2
+ p_\mu p^\mu\right),
\end{equation}
where $\mu = 0, 1, 2, 3$ and $p_\mu$ is the momentum conjugate to the
coordinate $x^\mu$ and $x^0$ is the time coordinate ($x^0 = t$). The
super-Hamiltonian constraint ${\cal H} = 0$ implies
\begin{equation}
\label{2}
m^2 + p^2 = m^2 + p_\mu p^\mu = 0.
\end{equation}
The Hamilton equations
\begin{eqnarray}
\label{3}
\dot x^\mu & =
{\partial{\cal H}\over\partial p_\mu} = {p^\mu\over m}, \hbox{\ \ and} \\
\dot p_\mu & =-{\partial{\cal H}\over\partial x^\mu}=0, \nonumber
\end{eqnarray}
where the dot means the derivative with respect to the affine parameter
$\lambda$, provide the expression for the momenta
\begin{equation}
\label{4}
p^\mu = m \dot x^\mu,
\end{equation}
and the second order equations of motion
\begin{equation}
\label{5}
\ddot x^\mu = 0.
\end{equation}
The super-Hamiltonian constraint provides a relation between the
affine parameter $\lambda$ and the proper time $\tau$ along the world
line of the particle.

The Lagrangian is related to the super-Hamiltonian ${\cal H}$ via
\begin{equation}
\label{6}
{\cal H} = p_\mu \dot x^\mu - L,
\end{equation}
and can be explicitly obtained in the following simple way:
\begin{equation}
\label{7}
L = p_\mu \dot x^\mu - {\cal H} = m \dot x^\mu \dot x^\mu -
{m\over 2} \left( 1 + \dot x^\mu \dot x_\mu\right) = {m\over 2}
\left( \dot x_\mu \dot x^\mu - 1\right).
\end{equation}
After separating the dynamical variables $x^i, i=1, 2, 3$ from the external
parameter $t=x^0$, we obtain an expression for the
Lagrangian.
\begin{equation}
\label{8}
L = {m\over 2} \left( \dot x_i \dot x^i - \dot t^2 - 1\right).
\end{equation}
The standard definition of the momenta conjugate to the dynamical degrees
of freedom $x^i$,
\begin{equation}
\label{9}
p_i = {\partial L\over\partial\dot x^i} = m \dot x_i,
\end{equation}
reproduces three (not four) Hamilton equations.
\begin{equation}
\label{10}
\dot x_i = {p_i\over m}.
\end{equation}
Using them, we construct the Hamiltonian $H$ of our theory,
\begin{equation}
\label{11}
H = p_i \dot x^i - L = {1\over m} p_i p^i - {1\over 2m} \left( p_i p^i -
m^2 \dot t^2 - m^2 \right),
\end{equation}
which, after simplifications, leads to
\begin{equation}
\label{12}
H = {1\over 2m} \left( p_i p^i + m^2 + m^2 \dot t^2\right).
\end{equation}
The Hamilton equations are obtained as usual, i.e.
\begin{eqnarray}
\label{13}
\dot x^i &= {\partial H\over\partial p_i } = {p_i\over m}, \nonumber \\
\dot p_i &= -{\partial H\over\partial x^i} = 0.
\end{eqnarray}
The super-Hamiltonian constraint is imposed in addition to the Hamilton
equations,
\begin{equation}
\label{14}
p_i p^i - m^2 \dot t^2 = - m^2,
\end{equation}
and completes the system of equation describing the particle motion.
It can be written (using Hamilton equations) as
\begin{equation}
\label{15}
dx_i dx^i - dt^2 = -d\lambda^2.
\end{equation}
The parameter $\lambda$ again coincides with the proper time
on the particle world line.

Although our new Hamiltonian $H$ of a free particle is
conserved, i.~e.~it is an integral of motion
\begin{equation}
\label{16}
{d H\over d \lambda} = m \dot t \ddot t = 0,
\end{equation}
it is, nevertheless, time dependent.
\begin{equation}
\label{17}
H = H(x^i, p_i, t(\lambda )).
\end{equation}
The Hamilton--Jacobi equation is obtained from this Hamiltonian.
\begin{equation}
\label{18}
{\partial S\over\partial\lambda} = - H\left( x^i, {\partial S\over \partial
x^i}, t(\lambda )\right).
\end{equation}

For this we use the expression \ref{12} for the Hamiltonian rewritten in the
following form:
\begin{equation}
\label{19}
H = {1\over 2m} g^{ik} p_i p_k + {m\over 2} (1 + \dot t^2);
\end{equation}
which leads to the Hamilton--Jacobi equation.
\begin{equation}
\label{20}
{\partial S\over \partial \lambda } = -{1\over 2m} g^{ik}
{\partial S \over \partial x^i} {\partial S \over \partial x^k} -
{m\over 2} (1 + \dot t^2).
\end{equation}
The Hamilton--Jacobi equation \ref{20} does not provide a complete description
of the particle dynamics as was the case for the Hamilton equations
\ref{13}. It too should be augmented by the constraint \ref{14} in order to
recover general Lorentz covariance.

The standard prescription of transition from the Hamilton--Jacobi equation to
the Schr\"odinger equation,
\begin{equation}
\label{21}
{\partial S\over\partial \lambda} \longrightarrow i \hbar {\partial \over
\partial \lambda}; \qquad
{\partial S\over\partial x^i} \longrightarrow \widehat p_i = {\hbar\over i}
{\partial \over \partial x^i },
\end{equation}
leads to the Schr\"odinger equation of a relativistic free particle,
\begin{equation}
\label{22}
i \hbar {\partial\psi\over\partial \lambda} =
{\hbar^2\over 2m}\, \Delta\psi - {m\over 2} (1 + \dot t^2)\psi.
\end{equation}
Here $\psi = \psi (x^i, \lambda )$ is the wave function of the free
particle and $\Delta = g^{ik} {\partial \over \partial x^i} {\partial
\over \partial x^k}$ is the Laplacian. The equation is a linear
differential equation with variable coefficients. It supports the
superposition principle and can be solved using the standard
techniques of quantum mechanics.  Moreover, it can be interpreted as
the Schr\"odinger equation for a particle in an external
spatially-homogeneous but time-dependent potential field.  This
time--dependent field is determined by the second term of the right
hand side ($\dot t$ is assumed to be a function of $\lambda$ only) of
\ref{22}.

As usual, the general solution of the equation \ref{22} depends on a
constant of integration (it is determined by the condition of
normalization of the $\psi$--function) as well as the initial
conditions (those, roughly speaking, specify the initial shape of the
wave packet). An unusual feature of the solution is its dependence on
a potential term that has not been specified yet. This feature is
related to the fact that in our formulation of the classical theory,
the Hamilton--Jacobi equation does not provide a complete dynamic
picture.  To complete the dynamic picture we need to use an analog of
the constraint \ref{14}. Our proposal is to use literally the same
constraint substituting in it instead of the classical values of
momenta $p_i$ their expectation values,
\begin{equation}
\label{23}
 <p_i (\lambda )> =
\langle\psi\vert\widehat p_i\vert\psi\rangle =
\int \psi^*\left( x^i, t(\lambda )\right)\, \widehat p_i\, \psi \left( x^i,
t(\lambda )\right)\, d^3x.
\end{equation}
The Schr\"odinger equation \ref{22} together with the constraint
\ref{14} obviously provide a complete dynamic description of the free
relativistic particle.

It is clear that the procedure of quantization described here avoids
the square--root Hamiltonian problem on both the quantum and classical
levels. This problem on the quantum level disappears due to the
elimination of $t$ from the set of dynamical variables.  On the
classical level the constraint equation \ref{14} is solved for $\dot
t$ which is a square root of nonnegative expression; therefore, it
avoids such square-root operators. The parameter $\lambda$ in the
quantum mechanical picture can be interpreted as the proper time of an
averaged quantum--driven distribution ``classical'' observer. The last
sentence should not be interpreted too literally. Ordinarily, the
observer is not classical as his motion is not described by the
classical equations of motion.\cite{HKM94b}

\section{Quantum Geometrodynamics of the Bianchi 1A Cosmology.}
\label{IV}

The Bianchi 1A cosmological model is commonly referred to as the axisymmetric
Kasner model. Its metric is determined by two parameters, the scale factor
$\Omega$ and the anisotropy parameter $\beta$.
\begin{equation}
\label{24}
ds^2 = - dt^2 + {\rm e}^{-2\Omega} \left( {\rm e}^{2\beta} dx^2 +
{\rm e}^{2\beta} dy^2 + {\rm e}^{-4\beta} dy^2 \right).
\end{equation}
As this cosmology is homogeneous the two functions $\Omega$ and $\beta$ are
the functions of the time parameter $t$ only. The scalar 4--curvature can
be expressed in terms of these two functions to yield the Hilbert action and,
after subtracting  the boundary term, the cosmological action,
\begin{equation}
\label{25}
I_C = I_H + {3V\over 8\pi}\dot\Omega {\rm e}^{-3\Omega}\vert_{t_0}^{t_f} =
{3V\over 8\pi}\int\limits_{t_0}^{t_f} \left( \dot\beta^2 - \dot\Omega^2
\right) {\rm e}^{-3\Omega} dt,
\end{equation}
where $V = \int\int\int dx dy dz$ is the spatial volume element.

We treat the scale factor $\Omega (t)$ as the many-fingered time parameter and
the anisotropy $\beta (t)$ as the dynamical degree of freedom. The momentum
conjugate to $\beta$ is
\begin{equation}
\label{26}
p_\beta = {\partial L \over \partial\dot\beta} = {3V\over 4\pi}
{\rm e}^{-3\Omega} \dot\beta.
\end{equation}
The Hamiltonian of the system in our approach can be expressed in terms of
the momentum conjugate to $\beta$ and the Lagrangian.
\begin{eqnarray}
\label{27}
H & = & p_\beta \dot\beta - L \nonumber \\
  & = & {4\pi \over 3V} {\rm e}^{3\Omega} p_\beta^2 - {3V \over 8\pi} \left(
        {4\pi \over 3V}\right)^2 {\rm e}^{3\Omega} p_\beta^2 + {3V\over 8\pi}
        \dot\Omega^2 {\rm e}^{-3\Omega} \nonumber \\
  & = & {2\pi \over 3V} {\rm e}^{3\Omega} p_\beta^2 + {3V\over 8\pi}
\dot\Omega^2 {\rm e}^{-3\Omega}.
\end{eqnarray}
In the classical theory this Hamiltonian can be used to produce either one
pair of Hamilton equations or the equivalent Hamilton--Jacobi
equation.  In any case, the dynamical picture derived in this way is
incomplete. To complete it we impose the super-Hamiltonian
constraint.
\begin{equation}
\label{28}
p_\beta^2 = \left(  {3V \over 4\pi}\right)^2 {\rm e}^{-6\Omega} \dot\Omega^2.
\end{equation}

Using the Hamilton--Jacobi equation,
\begin{equation}
\label{29}
{\partial S\over \partial t} = - H\left( {\partial S\over \partial\beta},
\Omega (t), \dot\Omega (t) \right),
\end{equation}
together with the expression \ref{27} for the Hamiltonian $H$ and the standard
quantization prescription we obtain the Schr\"odinger equation for
the axisymmetric Kasner model.
\begin{equation}
\label{30}
i\hbar {\partial\psi\over\partial t} = -{2\pi\hbar^2 \over 3V}
{\rm e}^{3\Omega} {\partial^2 \psi \over \partial\beta^2} +
{3V\over 8\pi} \dot\Omega^2 {\rm e}^{-3\Omega} \psi.
\end{equation}
The constant $\hbar$ in this equation should be understood as the square of
Planck's length scale, rather than the standard Planck constant. We wish to
stress here that the scale factor $\Omega$ in the Schr\"odinger equation is
so far an unknown function of time. This means that the equation does not
describe completely the quantum dynamics of the axisymmetric Kasner model.
To complete the dynamical picture we follow our prescription and impose, in
addition to equation \ref{30}, the super-Hamiltonian constraint.
\begin{equation}
\label{31}
<p_\beta>^2 = \left( {4\pi \over 3V}\right)^2 {\rm e}^{-6\Omega} \dot\Omega^2.
\end{equation}
Here $<p_\beta>$ is the expectation value of the momentum $\widehat p_\beta =
{\hbar\over i} {\partial \over \partial\beta}$
\begin{equation}
\label{32}
<p_\beta> = \langle\psi\vert\widehat p_\beta\vert\psi\rangle =
\int\limits_{-\infty}^{\infty} \psi^*(\beta , t) \widehat p_\beta \psi
(\beta , t) d\beta
\end{equation}
The system of equations \ref{30}, \ref{31} provide us with a complete
quantum dynamical picture of the axisymmetric Kasner model and, when
augmented by appropriate initial and boundary conditions, can be
solved analytically.

Before discussing the initial value conditions we will find the general
solution of the Schr\"odinger equation considering the scale factor $\Omega$
as a function of time generating an external potential. For this we separate
variables,
\begin{equation}
\label{33}
\psi (\beta , t) = \phi (\beta ) T(t).
\end{equation}
After substituting \ref{33} in the Schr\"odinger equation \ref{30} we obtain,
\begin{equation}
\label{34}
i\hbar \phi \dot T = -{2\pi\hbar^2\over 3V} {\rm e}^{3\Omega} T {\phi}'' +
{3 V\over 8\pi}\dot\Omega^2 {\rm e}^{-3\Omega} T \phi,
\end{equation}
where the prime means differentiation with respect to $\beta$. Rewriting it as
\begin{equation}
\label{35}
{2\pi\hbar^2\over 3V} {\phi''\over\phi} = -i\hbar {\rm e}^{-3\Omega}
{\dot T\over T} + {3V\over 8\pi} {\rm e}^{-6\Omega}\dot\Omega^2 = -\lambda,
\end{equation}
where $\lambda$ is the constant of separation, we obtain the equations for
$\phi (\beta )$ and $T(t)$.
\begin{equation}
\label{36}
\phi'' + {3V\over 2\pi\hbar^2} \lambda \phi = 0
\end{equation}
\begin{equation}
\label{37}
{\dot T\over T} = -{i\over\hbar} {\rm e}^{3\Omega} \left( {3V\over 8\pi}
{\rm e}^{-6\Omega} \dot\Omega^2 + \lambda \right)
\end{equation}
Equation \ref{36} admits only positive eigenvalues for $\lambda$.
Introducing the notation ${3V\lambda\over 2\pi} = k^2$ we can write
the solutions $\phi_k(\beta )$, $T_k(t)$ for $k \in (-\infty , \infty
)$.
\begin{eqnarray}
\label{38}
\phi_k(\beta) & = & A_k {\rm e}^{{i\over\hbar}k\beta} \nonumber \\
T_k(t) &  =  & B_k \exp\left\{ {-{i\over\hbar}\int_{t_0}^t \left( {2\pi\over
               3V} k^2 +{3V\over 8\pi} {\rm e}^{-6\Omega}\dot\Omega^2 \right)
               {\rm e}^{3\Omega} dt}\right\}
\end{eqnarray}
Using the superposition of these solutions we come up with the general
solution of the Schr\"odinger equation \ref{30}.
\begin{equation}
\label{39}
\psi (\beta, t) = \int\limits_{-\infty}^{\infty} A_k
{\rm e}^{{i\over\hbar}k\beta}
\exp\left\{ {-{i\over\hbar}\int_{t_0}^t \left( {2\pi\over
3V} k^2 +{3V\over 8\pi} {\rm e}^{-6\Omega}\dot\Omega^2 \right)
{\rm e}^{3\Omega} dt}\right\} dk
\end{equation}

To specify a particular problem one has to furnish appropriate initial data.
\begin{equation}
\label{40}
\psi (\beta , t)\vert_{t_0} = \psi (\beta , t_0) =
\int\limits_{-\infty}^{\infty} A_k {\rm e}^{{i\over\hbar}k\beta} dk
\end{equation}
It can be done either by specifying a function $\psi(\beta,t_0)$ and then
recovering $A_k$ from the equation
\begin{equation}
\label{41}
\psi (\beta , t_0) =
\int\limits_{-\infty}^{\infty} A_k {\rm e}^{{i\over\hbar}k\beta} dk,
\end{equation}
using Fourier transforms, or by assigning $A_k$ as a function of $k$,
depending on the type of the problem to be formulated. In this section we
consider the simplest example comparable with the quantum
mechanics of a particle, namely a wave packet. To describe a gaussian wave
packet centered initially at the value $k_0$ of $k$ (we will describe the
meaning of $k_0$ later) we assign
\begin{equation}
\label{42}
A_k = C {\rm e}^{-a(k - k_0)^2},
\end{equation}
where $C$ is the normalization constant. This leads to the following
expression for the initial values of the wave function:
\begin{equation}
\label{43}
\psi (\beta ,t_0) = C \int\limits_{-\infty}^{\infty} {\rm e}^{-a(k - k_0)^2}
{\rm e}^{{i\over\hbar}k\beta} dk = C \sqrt{{\pi\over a}}
{\rm e}^{{i\over\hbar}\beta k_0} {\rm e}^{-{\beta^2\over 4a\hbar^2}}.
\end{equation}
The value of the normalization constant $C$ is determined by the condition
\begin{equation}
\label{44}
\langle\psi\vert\psi\rangle = C^2 {\pi\over a} \int\limits_{-\infty}^{\infty}
{\rm e}^{-{\beta^2\over 2 a \hbar^2}} d\beta = C^2\hbar \pi^{{3\over 2}}
\sqrt{{2\over a}} = 1
\end{equation}
which leads to the value of $C^2$
\begin{equation}
\label{45}
C^2 = {\sqrt{a} \over \hbar\pi^{{3\over 2}}\sqrt{2}}.
\end{equation}
Using expression \ref{42} for $A_k$ and introducing notations for f and g,
\begin{eqnarray}
\label{46}
f = f(t) = {2\pi\over 3V}\int\limits_{t_0}^t {\rm e}^{3\Omega} dt, \nonumber \\
g = g(t) = {3V\over 8\pi} \int\limits_{t_0}^t \dot\Omega^2
                    {\rm e}^{-3\Omega} dt,
\end{eqnarray}
we can write down the solution $\psi (\beta, t)$ for the wave packet.
\begin{equation}
\label{47}
\psi (\beta, t) = C {\rm e}^{-{i\over\hbar}g} \int\limits_{-\infty}^{\infty}
{\rm e}^{-a(k -k_0)^2} {\rm e}^{{i\over\hbar}\beta k} {\rm e}^{{i\over\hbar}
f k^2} dk.
\end{equation}
After a simple transformation this expression can be rewritten in the form,
\begin{equation}
\label{48}
\psi (\beta ,t) = C \exp\left\{{i\over\hbar} \left[ (\beta - k_0 f) k_0 -
g\right]\right\} \int\limits_{-\infty}^{\infty} {\rm e}^{-ak^2}
{\rm e}^{-{i\over\hbar}fk^2} {\rm e}^{{i\over\hbar}(\beta - 2k_0f)k} dk.
\end{equation}
The integral on the right hand side of \ref{48} can be evaluated. The final
expression for the solution describing a gaussian wave packet may be
written in the following form which will prove to be convenient for
future calculations:
\begin{eqnarray}
\label{49}
\psi (\beta ,t)=C \sqrt{\pi} \left( a^2 + {f^2\over\hbar^2}
\right)^{-{1\over 4}}
\cdot \exp\left\{ -{a\over 4\left( a^2 + {f^2\over\hbar^2}\right)}
{(\beta - 2k_0f)^2\over\hbar^2} \right\} \times \nonumber \\
\exp\left\{ {i\over\hbar} (\beta - k_0f) k_0 \right\}
\cdot \exp\left\{ i{{f\over\hbar} \over 4\left( a^2 + {f^2\over\hbar^2}\right)}
{(\beta - 2k_0f)^2\over\hbar^2} \right\} \cdot \exp\left\{ -{i\over\hbar}g -
i \theta \right\};
\end{eqnarray}
where,
\begin{equation}
\cos ({2\theta}) = a/\sqrt{a^2 + f^2/\hbar^2}, \qquad
\sin ({2\theta}) = (f/\hbar)/\sqrt{a^2 + f^2/\hbar^2} \nonumber
\end{equation}
Although expression \ref{49} looks quite involved the last three
exponential factors are phase factors and do not complicate the
determination of the expectation values of the observables.

It is clear that this solution of the Schr\"odinger equation
describing the wave packet cannot provide any definite predictions as
it contains the two functions of time $f(t)$ and $g(t)$ which are
themselves related to the as yet undetermined scale factor $\Omega$.
To find $\Omega(t)$ we need to (1) compute the expectation $<p_\beta
>$ of the momentum $\widehat p_\beta = {\hbar\over i} {\partial
\over\partial\beta}$, (2)  substitute this expectation value into
the constraint \ref{31} and  (3)
solve the resulting equation with respect to $\Omega$. We start from
computing $<p_\beta >$.
\begin{equation}
\label{50}
<p_\beta > = \langle\psi\vert\widehat p_\beta \vert\psi\rangle = \\
C^2 \pi \left( a^2 + {f^2\over\hbar^2}\right)^{-{1\over 2}} k_0
\int\limits_{-\infty}^{\infty} \exp \left\{ -{a\over 2\left( a^2 +
{f^2\over\hbar^2}\right)}
{(\beta - 2k_0f)^2\over\hbar^2} \right\} d\beta = k_0.
\end{equation}
In other words the expectation value of the momentum $<p_\beta >$ does not
change with time. It is determined by the $k$--center of the packet at
$t=t_0$. Substitution of this result in \ref{31} yields
\begin{equation}
\label{51}
k_0^2 = \left( {3V \over 4\pi}\right)^2 {\rm e}^{-6\Omega} \dot\Omega^2.
\end{equation}
This equation and the classical equations are identical. Therefore,
we need not describe it in detail. We only wish to point out once more
that after the solution of this equation is substituted in \ref{49}
the geometrodynamic problem \ref{30}, \ref{31} for the wave packet
\ref{42} is solved completely. To summarize, the many--fingered time
of quantum geometrodynamics in case of a gaussian wave packet of
axisymmetric Kasner spacetimes coincides with its classical
counterpart if the expectation value of the momentum of the packet is
identified with the (conserved) value of the momentum of the classical
solution.

The expectation value for the anisotropy parameter $\beta$, where
$\beta$ is the only quantum dynamical variable in this model, is given
by:
\begin{equation}
\label{52}
<\beta > = \langle\psi\vert\beta\vert\psi\rangle = \\
C^2\pi \left(a^2 + {f^2\over\hbar^2}\right)^{-{1\over 2}}
\int\limits_{-\infty}^{\infty} \beta \exp\left\{ -{a\over 2\left( a^2 +
{f^2\over\hbar^2}\right)} {(\beta - 2k_0f)^2\over \hbar^2} \right\} d\beta
= 2k_0f(t).
\end{equation}
Thus ``the center'' of the wave packet evolves as the classical Kasner
universe determined by the momentum value equal to $k_0$ would evolve.
The spread of the wave packet with time is the variance in $\beta$.
\begin{eqnarray}
\label{53}
<\left(\beta\  - <\beta >\right)^2> = \nonumber \\
C^2\pi \left(a^2 + {f^2\over\hbar^2}\right)^{-{1\over 2}}
\int\limits_{-\infty}^{\infty} (\beta - 2k_0f)^2 \exp\left\{ -{a\over 2\left(
a^2 + {f^2\over\hbar^2}\right)} {(\beta - 2k_0f)^2\over \hbar^2} \right\}
d\beta =
{\hbar^2 a^2 + f^2 \over a}
\end{eqnarray}
It is obvious from \ref{53} that the spread of the packet increases
with time.  The result is similar to that of the quantum mechanics of
a free particle; after all the Bianchi~I cosmology is the
free--particle analogue of quantum cosmology.

\section{Characteristic Features of the Quantum Geometrodynamic Procedure.}
\label{V}

The quantum dynamical picture described in the previous sections has
two features that are not encountered typically in most of the more
common quantum dynamical schemes. They should be kept in mind in order
to avoid errors and misinterpretations in applying our procedure.

The first such feature is related to the structure of the Hamiltonian.
Formally, the Hamiltonian appears as a Hamiltonian of a system placed
in external time-dependent field, at least when the Schr\"odinger
equation is analyzed. However, the external field is determined via
constraints by the quantum state of the system.  Ordinarily, the
Hamiltonian depends on the initial data.  This feature is not
unique for our approach as it is encountered in standard quantum
mechanics such as Hartree--Fock--like systems.

The second feature is a split of the system parameters in two groups:
(1) the truly dynamical variables (to be quantized); and (2) the
descriptors of a ``natural observer'' associated to the system, driven
as they are by the quantum geometrodynamics. Such a split is
introduced prior to quantization, and is crucial to our theory.  This
feature can be clearly observed in both our examples. Essentially,
this split is a part of our solution of the time problem. Most of the
difficulties related to the problem of time are caused by an attempt
to enforce general covariance at the quantum level. Such an attempt
for gravitational systems is bound to fail. It leads to the questions
that are not well defined unless they are referred to a background
spacetime. Standard attempts to introduce a background spacetime tend
to use a fixed spacetime with properties generally not related to the
properties of the system itself, which is commonly considered as an
unsatisfactory feature (we quite agree with this conclusion). Our
procedure, on the other hand, can be considered as a universal
prescription for defining a unique background spacetime structure
driven by the quantum system. All questions concerning covariance
should be referred to this spacetime structure, and this spacetime
structure is ordinarily not classical. It cannot be considered as a
result of a 3--geometry evolution (the evolution equations do not take
a part in our procedure). In particular, we recover the classical
evolution only through an appropriately--peaked wave function and only
through constructive interference over conformal superspace.

Not every assignment of dynamical variables leads to a consistent
quantum dynamical picture. For example, a bad choice of the dynamical
component of the 3--geometry might lead to a nonelliptic differential
operator on the right hand side of Schr\"odinger equation.
Furthermore, under such conditions not every shape of the initial wave
packet will agree with the constraints. In any case, only an
appropriate choice of the set of dynamical variables will lead to a
reasonable quantum mechanical picture. The properties of the ADM
procedure together with the results of J.~York indicate that there is
at least one such reasonable choice. Is more than one possible quantum
dynamical picture for the same system, and if so, whether different
dynamical pictures lead to the same predictions?  This is a difficult
question to answer, notwithstanding the fact that we have been unable
to even properly formulate this question, as it is not only the
question of rewriting the equations but also an appropriate
reformulation of the initial conditions.  The initial conditions are
inextricably intertwined with the quantum dynamics.

\section{Discussion.}
\label{VI}

We have introduced in this paper a new approach for the quantization
of constrained systems and illustrated its application to two
examples. When the Bianchi~1A cosmology was quantized, our theory
generated a quantum geometrodynamic picture based on a post--ADM
treatment of the gravitational field dynamics and was free from the
conceptual difficulties usually associated with the Dirac and ADM
procedure of quantization. The variables describing the gravity field
in this case have been split into the true dynamical variables and a
parameter related to Wheeler's many-fingered time. Only the dynamical
part, the underlying anisotropy, has been quantized. The Hamiltonian
participating in the Schr\"odinger equation is not a square--root
Hamiltonian. This absence of a square--root Hamiltonian is generic for
our quantum geometrodynamic procedure.  The effective ``background
spacetime'' determined by the expectation values of dynamic variables
together with the ``observers'' (related to Wheeler's many--fingered
time) allows us to pose unambiguously the questions of covariance,
which in turn eliminates the problem of time in quantum gravity.
Furthermore, the Hilbert space problem does not appear to be the
generic feature in our view of quantum geometrodynamics.

The problems outlined in Sec.\ref{I} become all but eliminated by our
quantum geometrodynamic approach.  The nontrivial part of gravity
quantization appears to shift from such conceptual problems too the
problem of (1) the choice of an appropriate model to quantize, and to
the related problem of (2) the choice of an appropriate initial
condition for the wave functional.  Both choices are crucial if one is
to attempt using quantum geometrodynamics to better comprehend the
properties of gravitational systems. It is our understanding that the
success of gravity quantization rests on such meaningful choices.
Furthermore, the choice of models should not be determined by the
structure of quantum geometrodynamics; rather, it should be determined
by observational data and our general understanding of gravitational
phenomena.

It is clear that the procedure of quantizing the dynamical part of
constrained systems described in this paper can be extended to the
general case of geometrodynamics without any complications in
principle, although it may become quite involved computationally. The
procedure differs only in two respects from the simplistic examples
presented here. The first difference arises when the 3-metric is
(1) parametrized by three coordinatization parameters, (2) the
many--fingered time parameter, and (3) the two dynamical variables;
then all four constraints should be solved with respect to the
coordinatization parameters and the scale factor. In all four
constraints the expectation values of the true dynamic variables
should be used. The second difference is caused by the functional
nature of the gravitational field dynamics in the general case.  The
operation of functional integration is involved, which might lead to
analytic difficulties. Such difficulties are not specific for our
approach as they are common for the canonical formulations of all
field theories. Quantum geometrodynamics, in particular, does not seem
to generate any specific new difficulties.  In this paper we forwarded
the beginnings of a quantization scheme consistent with York's
analysis of the gravitational degrees of freedom. Although we parallel
the original motivations of Misner and the ADM quantization procedure,
our particular imposition of the four constraint equations leads to a
weaker theory that in turn avoids the problem of time.  As a more
mathematically--intricate description of our theory would be beyond
the scope of this introductory paper, we will publish the mathematical
foundations in a separate paper.\cite{HKM94}

We have not discussed here the inclusion of matter into quantum
geometrodynamics.  One simply should follow the pattern of matter
inclusion in the initial--value problem as outlined by J.~York.
However, this inclusion could be important, especially in the cases
where the matter degrees of freedom are coupled with the true
dynamical gravitational degrees of freedom${}^{11}$. Nevertheless, we
do not include it in the present paper, as it does not contribute to
the clarity of presentation of our ideas concerning the quantization
of constrained system in general, and quantum geometrodynamics in
particular.

We have demonstrated that the concept of time is inextricably
intertwined and woven to the initial conditions as well as to the
quantum dynamics over the space of all conformal 3-geometries.

\acknowledgements

For discussion, advice, or judgment on one or another issue taken up
in this manuscript, we are indebted R.~Fulp, S.~Habib, D.~Holz, R.
Laflamme, R.~Matzner, L.~Shepley, and J.~A.~Wheeler.

\end{document}